# Estimating the Lengths of Memory Words

Gusztáv Morvai, Benjamin Weiss

*Abstract*—For a stationary stochastic process $\{X_n\}$ with values in some set $A$, a finite word $w \in A^K$ is called a memory word if the conditional probability of $X_0$ given the past is constant on the cylinder set defined by $X_{-K}^{-1} = w$. It is a called a minimal memory word if no proper suffix of $w$ is also a memory word. For example in a $K$-step Markov processes all words of length $K$ are memory words but not necessarily minimal. We consider the problem of determining the lengths of the longest minimal memory words and the shortest memory words of an unknown process $\{X_n\}$ based on sequentially observing the outputs of a single sample $\{\xi_1, \xi_2, ...\xi_n\}$. We will give a universal estimator which converges almost surely to the length of the longest minimal memory word and show that no such universal estimator exists for the length of the shortest memory word. The alphabet $A$ may be finite or countable.

## I. INTRODUCTION

For a stationary stochastic process $\{X_n\}$ with values in some set $A$, a finite word $w \in A^K$ is called a memory word if for the conditional probability of $X_0$ given the past is constant on the cylinder set defined by $X_{-K}^{-1} = w$. We are using here the customary notation $\{X_i^j = X_i, X_{i+1}, ...X_j\}$. For example in a $K$-step Markov processes all words of length $K$ are memory words. However, in general $K$-step Markov processes may also have short memory words, cf. Bühlmann and Wyner [2]. Naturally any left extension of a memory word is also a memory word and it is natural to consider the minimal memory words, namely those none of whose proper suffixes are memory words. We consider the problem of determining the lengths of the longest minimal memory words and of the shortest memory words of an unknown process $\{X_n\}$ based on sequentially observing the outputs of a single sample $\{\xi_1, \xi_2, ...\xi_n\}$. That is to say we would like to have sequences of functions $L_n, S_n$ so that $L_n(\xi_1, \xi_2, ...\xi_n)$ will converge almost surely to the $K$ in case the process is $K$-step Markov (but not $(K-1)$-step Markov), and to infinity otherwise, while $S_n(\xi_1, \xi_2, ...\xi_n)$ will converge almost surely to the length of the shortest memory word in the process.

Most previous work of this kind (see for example Csiszár and Shields [3], Csiszár [4] and Ryabko et. al. [18]) was restricted to finite state processes. Our estimators will allow for countable alphabets and this precludes the use of a priori exponential estimates which can be used in the class of finite state $K$-step Markov chains. In the next section we will give

Gusztáv Morvai is with MTA-BME Stochastics Research Group, 1 Egry József utca , Building H, Budapest,1111, Hungary, e-mail: morvai@math.bme.hu, The first author was supported by the Bolyai János Research Scholarship. / Ez a cikk a Bolyai János Kutatási Ösztöndíj támogatásával készült.

Benjamin Weiss is with Hebrew University of Jerusalem, Jerusalem 91904 Israel, e-mail: weiss@math.huji.ac.il.

a universal estimator which converges almost surely to the length of the longest minimal memory word i.e. the minimal order of the process. This will be a finite number in case the process is a finite step Markov chain and infinity othewise. This is somewhat simpler than the estimators that we gave in [14]. On the other hand, we will show in the last section, that no sequence like $S_n$ which converges to the length of the shortest memory word can exist. In addition we will see how this gives a brief proof of another negative result of ours from [15] concerning estimators that are only defined along some sequence of stopping times.

## II. ESTIMATING THE LENGTH OF THE LONGEST MINIMAL MEMORY WORD OF A PROCESS

Let $\{X_n\}_{n=-\infty}^{\infty}$ be a stationary and ergodic time series taking values from a discrete (finite or countably infinite) alphabet $\mathcal{X}$. (Note that all stationary time series $\{X_n\}_{n=0}^{\infty}$ can be thought to be a two sided time series, that is, $\{X_n\}_{n=-\infty}^{\infty}$.) For notational convenience, let $X_m^n = (X_m, \ldots, X_n)$, where $m \leq n$. Note that if $m > n$ then $X_m^n$ is the empty string.

Let $p(x_{-k}^0)$ and $p(y|x_{-k}^0)$ denote the distribution $P(X_{-k}^0 = x_{-k}^0)$ and the conditional distribution $P(X_1 = y|X_{-k}^0 = x_{-k}^0)$, respectively.

*Definition 2.1:* We say that $w_{-k+1}^0$ is a memory word if $p(w_{-k+1}^0) > 0$ and for all $i \geq 1$, all $y \in \mathcal{X}$, all $z_{-k-i+1}^{-k} \in \mathcal{X}^i$

$$p(y|w_{-k+1}^0) = p(y|z_{-k-i+1}^{-k}, w_{-k+1}^0)$$

provided $p(z_{-k-i+1}^{-k}, w_{-k+1}^0, y) > 0$. If no proper suffix of $w$ is a memory word then $w$ is called a minimal memory word. Define the set $\mathcal{W}_k$ of those memory words $w_{-k+1}^0$ with length $k$, that is,

$$\mathcal{W}_k = \{w_{-k+1}^0 \in \mathcal{X}^k : w_{-k+1}^0 \text{ is a memory word}\}.$$

A discrete alphabet stationary time series is said to be a Markov chain if for some finite $K \geq 0$, $P(X_{-K}^0 \in \mathcal{W}_K) = 1$ and the smallest such $K$ is called the order of the Markov chain. For Markov chains the order is the length of the longest minimal memory word.

In general we can define a function $K$ which will give us the length of the minimal memory word of a sequence of past observations.

*Definition 2.2:* For a stationary time series $\{X_n\}$ the (random) length $K(X_{-\infty}^0)$ of the memory of the sample path $X_{-\infty}^0$ is the smallest possible $0 \leq K < \infty$ such that for all $i \geq 1$, all $y \in \mathcal{X}$, all $z_{-K-i+1}^{-K} \in \mathcal{X}^i$

$$p(y|X_{-K+1}^0) = p(y|z_{-K-i+1}^{-K}, X_{-K+1}^0)$$

provided $p(z_{-K-i+1}^{-K}, X_{-K+1}^0, y) > 0$, and $K(X_{-\infty}^0) = \infty$ if there is no such $K$.

Our goal in this section is to estimate the essential supremum of this function. This is a simple numerical function which depends on the process and not on any particular realization of the process. In contrast, in [15], we addressed the problem of estimating the minimal length of a memory word which occurs as the suffix of the first $n$ observations of the process. This varies of course with $n$ and with the realization. For $k \geq 0$ let $\mathcal{S}_k$ denote the support of the distribution of $X_{-k}^0$:
$$\mathcal{S}_k = \{x_{-k}^0 \in \mathcal{X}^{k+1} : p(x_{-k}^0) > 0\}.$$

Define
$$\Delta_k = \sup_{1 \leq i} \sup_{(z_{-k-i+1}^0, x) \in \mathcal{S}_{k+i}} \left| p(x|z_{-k+1}^0) - p(x|z_{-k-i+1}^0) \right|.$$

If for some $k$, $\Delta_k = 0$ then the process is a Markov chain and the least such $k$ is the order of the chain. We need to define a statistic to estimate $\Delta_k$. To this end let
$$\hat{p}_n(x|z_{-k+1}^0) =$$
$$\frac{\left(\#\{k-1 \leq t \leq n-1 : X_{t-k+1}^{t+1} = (z_{-k+1}^0, x)\} - 1\right)^+}{\left(\#\{k-1 \leq t \leq n-1 : X_{t-k+1}^t = z_{-k+1}^0\} - 1\right)^+}.$$

where $0/0$ is defined as $0$. We subtract one for technical reasons which does not effect its properties we need here. These empirical distributions, as well as the sets we are about to introduce are functions of $X_0^n$, but we suppress this dependence to keep the notation manageable.

For a fixed $0 < \gamma < 1$ let $\mathcal{S}_k^n$ denote the set of strings with length $k+1$ which appear more than $n^{1-\gamma}$ times in $X_0^n$. That is,
$$\mathcal{S}_k^n = \{x_{-k}^0 \in \mathcal{X}^{k+1} : \#\{k \leq t \leq n : X_{t-k}^t = x_{-k}^0\} > n^{1-\gamma}\}.$$

These are the strings which occur sufficiently often so that we can rely on their empirical distribution.

Finally, define the empirical version of $\Delta_k$ as follows:
$$\hat{\Delta}_k^n = \max_{1 \leq i \leq n} \max_{(z_{-k-i+1}^0, x) \in \mathcal{S}_{k+i}^n} \left| \hat{p}_n(x|z_{-k+1}^0) - \hat{p}_n(x|z_{-k-i+1}^0) \right|$$

Let us agree by convention that if the smallest of the sets over which we are maximizing is empty then $\hat{\Delta}_k^n = 0$.

Observe, that by ergodicity, for any fixed $k$,
$$\liminf_{n \to \infty} \hat{\Delta}_k^n \geq \Delta_k \quad \text{almost surely.} \tag{1}$$

We define an estimate $\chi_n$ for the order from samples $X_0^n$ as follows. Let $0 < \beta < \frac{1-\gamma}{2}$ be arbitrary. Set $\chi_0 = 0$, and for $n \geq 1$ let $\chi_n$ be the smallest $0 \leq k_n < n$ such that $\hat{\Delta}_{k_n}^n \leq n^{-\beta}$ if there is such a $k$ and $n$ otherwise.

*Theorem 2.1:* For any ergodic, stationary process $\{X_n\}$ the sequence of estimators $\chi_n$ converges almost surely to the essential supremum of the memory function $K$.

PROOF:
If the process is a Markov chain, it is immediate that for all $k$ greater than or equal the order, $\Delta_k = 0$. For $k$ less than the order $\Delta_k > 0$. If the process is not a Markov chain with any finite order then $\Delta_k > 0$ for all $k$. Thus by (1) if the process is not Markov then $\chi_n \to \infty$ and if it is Markov then $\chi_n$ is greater than or equal to the order eventually almost surely. We have to show that if the process is a Markov chain of order $K$ then $\chi_n$ is eventually almost surely at most $K$.

Let us suppose that the process is indeed a Markov chain with order $K$. Recall the simple fact that the letters $u_i$ that follow the successive occurrences of a word $w$ with length $K$ are independent and identically distributed random variables (cf. Lemma 1 in Morvai and Weiss [15]). Since the alphabet may be infinite we can't take into consideration all possible words in our estimation of the undesirable event $(\chi_n > K)$. Instead we restrict attention to the words that actually occur and so we fix a location $(l-K, l)$ in the index set and then fix a word $w_{-K+1}^0$ that occurs there together with a particular state $x$ that follows it. The random times $l+\lambda^+$ and $l-\lambda^-$ are the other occurrences of this memory word in the process. Here is the formal definition. Set $\lambda_{l,K,0}^+ = 0$, $\lambda_{l,K,0}^- = 0$ and define

$$\lambda_{l,K,i}^+ = \lambda_{l,K,i-1}^+ +$$
$$\min\{t > 0 : X_{l+\lambda_{l,K,i-1}^+ - K+1+t}^{l+\lambda_{l,K,i-1}^+ + t} = X_{l+\lambda_{l,K,i-1}^+ - K+1}^{l+\lambda_{l,K,i-1}^+}\}$$

and

$$\lambda_{l,K,i}^- = \lambda_{l,K,i-1}^- +$$
$$\min\{t > 0 : X_{l-\lambda_{l,K,i-1}^- - K+1-t}^{l-\lambda_{l,K,i-1}^- - t} = X_{l-\lambda_{l,K,i-1}^- - K+1}^{l-\lambda_{l,K,i-1}^-}\}$$

Assume $w_{-K+1}^0$ is any word and $x$ is a letter. Then by Lemma 1 in Morvai and Weiss [15] for $i, j \geq 1$,

$$X_{l-\lambda_{l,K,i}^- + 1}, \ldots, X_{l-\lambda_{l,K,1}^- + 1}, X_{l+\lambda_{l,K,1}^+ + 1}, \ldots, X_{l+\lambda_{l,K,j}^+ + 1}$$

are conditionally independent and identically distributed random variables given $X_{l-K+1}^l = w_{-K+1}^0, X_{l+1} = x$, where the identical distribution is $p(\cdot|w_{-K+1}^0)$.

Letting $n \geq K$ we proceed to estimate the probability of the undesirable event $(\chi_n > K)$.

Observe that by our definition of $\chi_n$ we have

$$P(\chi_n > K) \leq P(\hat{\Delta}_K^n > n^{-\beta}) \leq \sum_{i=1}^n P(\max_{(z_{-K-i+1}^0, x) \in \mathcal{S}_{K+i}^n}$$
$$\left| \hat{p}_n(x|z_{-K+1}^0) - \hat{p}_n(x|z_{-K-i+1}^0) \right| > n^{-\beta})$$

where the second inequality follows from our assumption on the order of the chain.

The last term satisfies:

$$\sum_{i=1}^{n} P(\max_{(z^0_{-K-i+1},x)\in \mathcal{S}^n_{K+i}} \left|\hat{p}_n(x|z^0_{-K+1}) - \hat{p}_n(x|z^0_{-K-i+1})\right| > n^{-\beta})$$

$$\leq \sum_{i=1}^{n} P(\max_{(z^0_{-K-i+1},x)\in \mathcal{S}^n_{K+i}} \left|\hat{p}_n(x|z^0_{-K+1}) - p(x|z^0_{-K+1})\right| > 0.5n^{-\beta})$$

$$+ \sum_{i=1}^{n} P(\max_{(z^0_{-K-i+1},x)\in \mathcal{S}^n_{K+i}} \left|p(x|z^0_{-K-i+1},z^0_{-K+1}) - \hat{p}_n(x|z^0_{-K-i+1})\right| > 0.5n^{-\beta}) \quad (2)$$

We continue with the estimation of the last two summands. For a given $K-1 \leq l \leq n-1$ assume that $X^{l+1}_{l-K+1} = w^0_{-K+1}x$. By Lemma 4.1 (Hoeffding's inequality) in the Appendix for sums of bounded independent random variables implies

$$P\left(\left|\frac{\sum_{h=1}^{i} 1_{\{X_{l-\lambda^-_{l,K,h}+1}=x\}} + \sum_{h=1}^{j} 1_{\{X_{l+\lambda^+_{l,K,h}+1}=x\}}}{i+j}\right.\right.$$
$$\left.-p(x|w^0_{-K+1})\right|$$
$$\geq 0.5n^{-\beta} \mid X^{l+1}_{l-K+1} = w^0_{-K+1}x\right) \leq 2e^{-0.5n^{-2\beta}(i+j)}.$$

Multiplying both sides by $P(X^{l+1}_{l-K+1} = w^0_{-K+1}x)$ and summing over all possible words $w^0_{-K+1}$ and $x$ we get that

$$P\left(X^{l+1}_{l-K+1} \in \mathcal{S}^n_{K+1},\right.$$
$$\left|\frac{\sum_{h=1}^{i} 1_{\{X_{l-\lambda^-_{l,K,h}+1}=X_{l+1}\}} + \sum_{h=1}^{j} 1_{\{X_{l+\lambda^+_{l,K,h}+1}=X_{l+1}\}}}{i+j}\right.$$
$$\left.- p(X_{l+1}|X^l_{l-K+1})\right| > 0.5n^{-\beta})$$
$$\leq 2e^{-0.5n^{-2\beta}(i+j)}.$$

Summing over all $K-1 \leq l \leq n-1$ and over all pairs $(i,j)$ such that $i \geq 0$, $j \geq 0$, $i+j \geq \lfloor n^{1-\gamma} \rfloor$ we get that

$$P\left(\text{For some } K-1 \leq l \leq n-1 : X^{l+1}_{l-K+1} \in \mathcal{S}^n_{K+1},\right.$$
$$\left.\left|\hat{p}_n(X_{l+1}|X^l_{l-K+1}) - p(X_{l+1}|X^l_{l-K+1})\right| > 0.5n^{-\beta}\right)$$
$$\leq n^2 \sum_{h=\lfloor n^{1-\gamma} \rfloor}^{\infty} h 2e^{-0.5n^{-2\beta}h}.$$

Applying this final inequality to each of the terms in (2) we get that

$$P(\chi_n > K) \leq 4n^3 \sum_{h=\lfloor n^{1-\gamma} \rfloor}^{\infty} h e^{0.5n^{-2\beta}h}$$

The right hand side is summable given $0 < \beta < \frac{1-\gamma}{2}$ and then by the Borel-Cantelli lemma a.s. the undesirable event occurs only finitely many times and thus the proof of Theorem 2.1 is complete.

**Remark.** Bailey [1] showed that one can not discriminate between processes where the supremum of the lengths of the minimal memory words is finite or infinite. In other words, there is no sequence of functions which will converge to **Yes** in case the observed process is Markov with some finite but unknown order and to **No** otherwise. The result we have just established does not contradict this since our estimators give numbers rather than just two values. There is a more detailed discussion of this phenomenon in [12].

## III. LIMITATIONS ON ESTIMATING THE LENGTH OF THE SHORTEST MEMORY WORD OF A SECOND ORDER MARKOV CHAIN

The next theorem shows that even when we restrict attention to second order Markov chains there is no universal estimator for the length of the shortest memory word.

*Theorem 3.1:* Let $\mathcal{X} = \{0,1,2,\ldots\}$. For any estimator $\{h_n(X_0,\ldots,X_n)\}$ such that for all stationary and ergodic second order Markov chains taking values from $\mathcal{X}$ with minimum length of memory words being equal to one

$$\limsup_{n \to \infty} P(h_n(X_0,\ldots,X_n) = 1) = 1$$

there exists a stationary and ergodic second order Markov chain taking values from $\mathcal{X}$ with minimum length of memory words being equal to two such that

$$\limsup_{n \to \infty} P(h_n(X_0,\ldots,X_n) = 1) = 1.$$

PROOF: As is customary in proofs of this type of theorem we construct the problematic Markov chain by a sequence of steps in which at intermediate stages we will have a second order Markov chain with some memory words of length one. The point is that these memory word occur very infrequently so that a very small modification of the process suffices to destroy one of these while preserving the others. This modification is small enough so as not to change some finite distributions by too much so that all of these features will be present in the limiting process. To keep the technical details to a minimum we start with one Markov chain and all of our modification are functions of the initial chain. Here are the details.

First we define a Markov-chain (cf. Ryabko [17]) which serves as the technical tool for construction of our counterexample. Let the state space $S$ be the non-negative integers. From state 0 the process certainly passes to state 1 and then to state 2, at the following epoch. From each state $s \geq 2$, the Markov chain passes either to state 0 or to state $s+1$ with equal probabilities 0.5. This construction yields a stationary and ergodic Markov chain $\{M_i\}$ with stationary distribution

$$P(M=0) = P(M=1) = \frac{1}{4}$$

and

$$P(M=i) = \frac{1}{2^i} \text{ for } i \geq 2.$$

Now let $f^{(0)}(0) = f^{(0)}(1) = 0$ and for all $s \geq 2$ let $f^{(0)}(s) = s$. The process $\{X^{(0)}_i = f^{(0)}(M_i)\}$ is a stationary ergodic countable alphabet second order Markov-chain with minimum length of the memory words being equal to one. Set $N_0 = 1$. Let $n_0 > N_0$ be so large such that

$$P\left(h_{n_0}(X_0^{(0)},\ldots,X_{n_0}^{(0)}) = 1 | X_0^{(0)} = X_1^{(0)} = 0\right) > 1 - \frac{1}{2}.$$

Now, observe that
- for $N_0 < s$: $f^{(0)}(s) = s$ and $s$ is a memory word,
- $\{f^{(0)}(s) : s \leq N_0\} \bigcap \{f^{(0)}(s) : N_0 < s\} = \emptyset$
- for $0 \leq s \leq N_0$: $f^{(0)}(s)$ is not a memory word.

Now we define the function $f^{(1)}$. For all $0 \leq s \leq n_0$ and for all $2n_0 - N_0 + 1 \leq s$ define

$$f^{(1)}(s) = f^{(0)}(s)$$

and for all $n_0 < s \leq n_0 + (n_0 - N_0)$ let $f^{(0)}(s) = n_0 - (s - n_0) + 1$. The resulting process $\{X_i^{(1)} = f^{(1)}(M_i)\}$ is a stationary ergodic countable alphabet second order Markov-chain with minimum length of the memory words being equal to one and

$$P\left(h_{n_0}(X_0^{(1)},\ldots,X_{n_0}^{(1)}) = 1 | X_0^{(1)} = X_1^{(1)} = 0\right) > 1 - \frac{1}{2}.$$

Put $N_1 = 2n_0 - N_0$ and let $n_1 > N_1 + 1$ be so large that

$$P\left(h_{n_1-1}(X_0^{(1)},\ldots,X_{n_1-1}^{(1)}) = 1 | X_i^{(1)} = X_{i+1}^{(1)} = 0\right.$$
$$\left. \text{for some } -1 \leq i \leq 0\right) > 1 - \left(\frac{1}{2}\right)^2.$$

Indeed, there exists such an $n_1$ since by assumption,

$$\limsup_{n \to \infty} P\left(h_n(X_0^{(1)},\ldots,X_n^{(1)}) = 1\right) = 1.$$

Now, observe that
- for $N_1 < s$: $f^{(1)}(s) = s$ and $s$ is a memory word,
- $\{f^{(1)}(s) : s \leq N_1\} \bigcap \{f^{(1)}(s) : N_1 < s\} = \emptyset$
- for $0 \leq s \leq N_1$: $f^{(1)}(s)$ is not a memory word.

Now we define the function $f^{(j)}$ inductively. Assume we have already defined positive integers $N_k$, $n_k > N_k + k$ and functions $f^{(k)}$ for $0 \leq k \leq j-1$ with the following properties:
- each process $X_n^{(k)} = f^{(k)}(M_n)$ is a stationary and ergodic countable alphabet second order Markov chain with minimum length of the memory words being equal to one,
- for $0 \leq s \leq n_k$, $f^{(j-1)}(s) = f^{(k)}(s)$,
- $P\left(h_{n_k-k}(X_0^{(j-1)},\ldots,X_{n_k-k}^{(j-1)}) = 1 | X_i^{(j-1)} = X_{i+1}^{(j-1)} = 0 \text{ for some } -k \leq i \leq 0\right) > 1 - \left(\frac{1}{2}\right)^{k+1}$
- for $N_{j-1} < s$: $f^{(j-1)}(s) = s$ and $s$ is a memory word,
- $\{f^{(k)}(s) : s \leq N_k\} \bigcap \{f^{(k)}(s) : N_k < s\} = \emptyset$,
- for $0 \leq s \leq N_{j-1}$ $f^{(j-1)}(s)$ is not a memory word.

Now we define the function $f^{(j)}$. For all $0 \leq s \leq n_j$ and for all $2n_{j-1} - N_{j-1} + 1 \leq s$ define

$$f^{(j)}(s) = f^{(j-1)}(s)$$

and for all $n_{j-1} < s \leq n_{j-1} + (n_{j-1} - N_{j-1})$ let $f^{(j)}(s) = n_{j-1} - (s - n_{j-1}) + 1$.

The resulting process $\{X_i^{(j)} = f^{(j)}(M_i)\}$ is a stationary ergodic countable alphabet second order Markov-chain with minimum length of the memory words being equal to one and for all $0 \leq k < j$

$$P\left(h_{n_k-k}(X_0^{(j)},\ldots,X_{n_k-k}^{(j)}) = 1 | X_i^{(j)} = X_{i+1}^{(j)} = 0\right.$$
$$\left. \text{for some } -k \leq i \leq 0\right) > 1 - \left(\frac{1}{2}\right)^{k+1}.$$

Put $N_j = 2n_{j-1} - N_{j-1}$ and let $n_j > N_j + j$ be so large such that

$$P\left(h_{n_j-j}(X_0^{(j)},\ldots,X_{n_j-j}^{(j)}) = 1 | X_i^{(j)} = X_{i+1}^{(j)} = 0\right.$$
$$\left. \text{for some } -j \leq i \leq 0\right) > 1 - \left(\frac{1}{2}\right)^{j+1}.$$

For the function $f^{(j)}$ just defined,
- the process $f^{(j)}(M_n)$ is a stationary and ergodic countable alphabet second order Markov chain with minimum length of the memory words being equal to one,
- for $0 \leq k < j$ and $0 \leq s \leq n_k$: $f^{(j-1)}(s) = f^{(k)}(s)$,
- for $0 \leq k < j$:

$$P\left(h_{n_k-k}(X_0^{(j)},\ldots,X_{n_k-k}^{(j)}) = 1 | X_i^{(j)} = X_{i+1}^{(j)} = 0\right.$$
$$\left. \text{for some } -k \leq i \leq 0\right) > 1 - \left(\frac{1}{2}\right)^{k+1}$$

- for $N_j < s$: $f^{(j)}(s) = s$ and $s$ is a memory word,
- $\{f^{(j)}(s) : s \leq N_j\} \bigcap \{f^{(j)}(s) : N_j < s\} = \emptyset$,
- for $0 \leq s \leq N_j$ $f^{(j)}(s)$ is not a memory word.

Eventually, we defined a function $f(s) = \lim_{k \to \infty} f^k(s)$ and a process $X_n = f(M_n)$ which is a stationary ergodic countable alphabet second order Markov-chain with minimum length of the memory words being equal to TWO and for all $0 \leq k$

$$P\left(h_{n_k-k}(X_0,\ldots,X_{n_k-k}) = 1 | X_i = X_{i+1} = 0\right.$$
$$\left. \text{for some } -k \leq i \leq 0\right) > 1 - \left(\frac{1}{2}\right)^{k+1}.$$

Thus

$$P\left(h_{n_k-k}(X_0,\ldots,X_{n_k-k}) = 1\right) > \left(1 - \left(\frac{1}{2}\right)^{k+1}\right)$$
$$P\left(X_i = X_{i+1} = 0 \text{ for some } -k \leq i \leq 0\right).$$

Since

$$\lim_{k \to \infty} P\left(X_i = X_{i+1} = 0 \text{ for some } -k \leq i \leq 0\right) = 1$$

we get that

$$\limsup_{n \to \infty} P\left(h_n(X_0,\ldots,X_n) = 1\right) = 1.$$

The proof of Theorem 3.1 is complete.

The following theorem has been proved in Morvai and Weiss [15] (Theorem 6). Here we give a simpler proof of it based on Theorem 3.1.

*Theorem 3.2:* Let $\mathcal{X} = \{0, 1, 2, \ldots\}$. There are no strictly increasing sequence of stopping times $\{\lambda_n\}$ and estimators $\{h(X_0, \ldots, X_{\lambda_n})\}$ taking the values one and two, such that for all second order Markov chains taking values from $\mathcal{X}$:

$$\lim_{n \to \infty} \frac{\lambda_n}{n} = 1$$

and

$$\lim_{n \to \infty} |h(X_0, \ldots, X_{\lambda_n}) - K(X_0^{\lambda_n})| = 0 \text{ with probability one.}$$

PROOF:

We argue by contradiction. Assume that Theorem 3.2 does not hold. Then define

$$\hat{l}_n(X_0, \ldots, X_n) = \min_{0.5n < \lambda_k < n} h(X_0, \ldots, X_{\lambda_k}).$$

Now, by assumption $\hat{l}_n(X_0, \ldots, X_n)$ would be a pointwise consistent estimate for the length of the shortest memory word which contradicts Theorem 3.1. The proof of Theorem 3.2 is complete.

**Remark.** For a positive result using stopping times cf. Theorem 4 of [15]. It shows that for any positive $\epsilon$ there is a sequence of stopping times $\lambda_n$ which with probability one will have density at least $1 - \epsilon$ and along which we can successfully estimate $K(X_0^{\lambda_n})$. For more on the use of stopping times in universal estimation see the recent survey [16] and [7], [8], [9], [13], [10], [11].

## IV. APPENDIX

The next lemma is due to Hoeffding, cf. [6].

*Lemma 4.1:* (Hoeffding's inequality, Hoeffding 1963) Let $X_1, X_2, \ldots, X_n$ be independent real valued random variables, and $a_1, b_1, \ldots, a_n, b_n$ be real numbers such that $a_i \leq X_i \leq b_i$ with probability one for all $1 \leq i \leq n$. Then, for all $\epsilon > 0$,

$$P\left(\left|\frac{1}{n}\sum_{i=1}^{n}(X_i - EX_i)\right| > \epsilon\right) \leq 2e^{-\left(2n\epsilon^2 / \frac{1}{n}\sum_{i=1}^{n}|b_i - a_i|^2\right)}.$$